# Business and Information Technology Alignment Measurement - a recent Literature Review (Preprint version)

Leonardo Muñoz[1] and Oscar Avila[1]

[1] Department of Systems and Computing Engineering, School of Engineering,
Universidad de los Andes, Bogotá, Colombia
{l.munozm, oj.avila}@uniandes.edu.co

**Abstract.** Since technology has been involved in the business context, Business and Information Technology Alignment (BITA) has been one of the main concerns of IT and Business executives and directors due to its importance to overall company performance, especially today in the age of digital transformation. Several models and frameworks have been developed for BITA implementation and for measuring their level of success, each one with a different approach to this desired state. The BITA measurement is one of the main decision-making tools in the strategic domain of companies. In general, the classical-internal alignment is the most measured domain and the external environment evolution alignment is the least measured. This literature review aims to characterize and analyze current research on BITA measurement with a comprehensive view of the works published over the last 15 years to identify potential gaps and future areas of research in the field.

## 1 Introduction

Currently all organizations are leveraged by technology at different levels, from the operational to the strategical. Due to this growing relationship, several frameworks and approaches have been developed to model and achieve alignment between traditional business structures and the technology domain. The main objective of these Business and Information Technology Alignment (BITA) models and methods is to transform the way business and Information Technology (IT) domains understand each other in terms of objectives and requirements in business execution.

In this line of ideas, one of the biggest problems in organizations is the misalignment between business and IT objectives and needs. Because each of these domains works independently for their improvement through individual frameworks, the result is failed projects and delays in overall company performance. For this reason, BITA has been one of the main concerns of business and IT directors in the last decade [1-3], due to the fact that BITA helps to close the gaps in communication and interaction between Business and IT domains in organizations. This implies that also measuring and evaluating such alignment is an important concern.

The importance of BITA measurement in the organizations is that it constitutes one of the main sources of information for the decision-making process. Commonly, BITA measurement have been focused on the "classical" internal alignment, which is related to the measurement of the alignment between the business strategy and processes and the IT resources. But, the modern organizations must have to evaluate two other alignment levels that are, the alignment with the external environment and the alignment with uncertain future evolutions [4]. These alignments with the environment and future evolutions are not currently the focus on the BITA measurement area, what means that the organizations are not prepared to the modern enterprise environment which is in continuous and fast changing. This lack of focus on the alignment with future evolutions and external environment could be one of the main problems of organizations because they depend on the changing external market. The BITA measurement at these levels could improve the sustainability and growth of the companies.

Considering this concern, the main objective of this article is to evaluate the most recent works in the literature of BITA measurement to identify their contributions to the three alignment levels above mentioned: classical strategic alignment, alignment with the external environment and alignment with the external evolutions. Based on the evaluation results, our purpose is to present the potential gaps and propose future areas of research. This article is organised as follows: the second section describes the study methodology and evaluation framework to review existing literature, which includes evaluation categories, criteria and questions. The third section describes how the research literature was evaluated. The fourth section describes the application of the framework and the synthesis and analysis of corresponding results. Finally, in the fifth section, the findings and conclusions are presented to lay foundations for future research.

## 2      Study methodology and evaluation framework

The literature review process includes:

- *(i)*     *Planning.* It consists in the selection of the categories and criteria which will be used to design the evaluation framework for the selected articles.

- *(ii)*     *Realization.* This step consists in the definition of the research terms to make the search and selection of the articles to be evaluated.

(iii)   *Synthesis and analysis.* This step presents the evaluation framework application to the selected articles and the analysis of the results for each research question.

## 2.1   Planning

This work aims to evaluate the last 15 years of research literature on BITA measurement methods and approaches. For the evaluation of this set of articles, the proposed framework in [5] was adopted as a basis to develop a personalized evaluation framework. To classify the BITA approaches, the Strategic Alignment Model (SAM) [6] proposed by Henderson and Venkatraman was also adopted as one of the base criteria to evaluate the scope of the works being reviewed. Based on this, a characterization framework was defined to review current research on BITA measurement methods and approaches. This framework has three main categories, each with specific criteria for proposing research questions (See Table 1).

*Context.* This category looks for identifying the objective of the reviewed works and identifying common characteristics in such objetives.

*Alignment.* This category aims to understand the alignment scope of BITA measurement models in terms of the three alignment levels mentioned above (see introduction section). This analysis is intended to know at which alignment levels measurement is made at each approach. According to [4], in BITA we can classify the alignment in: First, the classical-internal alignment that aims to align all the organization areas with IT. Second, the alignment with the environment, referring to the external actors and events which could affect the organization. Finally, the alignment with uncertain evolutions related to future changes in the internal domains of the organization and the external environment.

*Measurement.* This category examines the nature, the type and other details of the measurement method. This is the focus of our work and for this reason it is the main section of the proposed framework in this article (see Table 1).

**Table 1.** Proposed evaluation framework.

| Category | Criteria | Questions |
|---|---|---|
| Context | Objective | What is the main objective of the article? |
| Alignment | Strategic alignment | Which organizational domains of the business context can be aligned? |
| | | According to the SAM alignment model, which is the alignment sequence of the applied model? |
| | Environment alignment | Which external environment elements of the business and IT are aligned? |
| | Future alignment | Which temporal dimensions are approached by the alignment model used? |
| | Alignment level | Which alignment levels are addressed? |

| Measurement | Method type | Which is the type of measurement method approached? |
|---|---|---|
| | Measurement Nature | Which is the nature of the measurement method approached? |
| | Metrics | Which are the measurement criteria and metrics of the method? |
| | Methodology | Which are the steps included in methodology applied? |

## 3 Realization

For conducting the identification of the related articles, SCOPUS was adopted as search tool, keeping in mind it is the largest and one of the main scientific abstract and citation databases [7].

To find the set of academic articles for the evaluation, the search query showed below was used on the SCOPUS platform:

*TITLE-ABS( "Alignment" OR "Match" OR "Fit" OR "Fitness" OR "Strategic alignment" OR "Coordination" OR "Link" ) AND TITLE-ABS( "Business and IT" OR "Business/IT" OR "Business-IT" OR "Business/IS" OR "Business and IS" OR "Business-IS" OR "Business and Information Systems" OR "Business and Information technology" OR "Business and Information technologies" ) AND TITLE-ABS( "Measurement" OR "Assessment" OR "Measure" OR "Assess" OR "Evaluation" OR "Evaluate" OR "measuring" OR "evaluating" OR "assessing" )*

The search yielded 386 results. A second filter was made by scanning the titles and abstracts to obtain a set of articles that were closely related to the measurement topic. Following, through the reading and detailed review, the final set was defined, consisting of 22 articles which present different methods and approaches for BITA measurement [8-29]. The proposed evaluation framework was applied to this set of articles, answering each of the research questions in each category and criteria. For the analysis of this set, reference works such as Henderson and Venkatraman and Luftman were used. These are fundamental for the BITA research and its measurement.

## 4 Synthesis and analysis

For the synthesis and analysis, we review the selected works with respect to each research question in the framework. In general, we found that the approaches in current literature are not homogenous. The synthesis of the review is presented below.

### 4.1 Context

**Which is the main objective of the research work?**

Across the 22 articles reviewed, we found that the focus in the last years of research was to present new methods and approaches for measuring alignment without a case study application (46% of the articles); articles that apply alignment methods to different specific cases of study (36% of the articles) and articles that present both, new approaches and its application (18% of the article). Many of the new proposed approaches have one or more previous frameworks as the base of development, producing thus enhanced methods through the adoption of one of the existing approaches or a mix of them.

## 4.2 Alignment

**What organizational domains of the business context can be aligned?**

Despite the variety of terminology, most of the reviewed works rely on to two main Business and IT modelling approaches:

First, the *EA metamodels* in [10-16], where layers, domains or other elements involved in alignment measuring are: Business architecture, data architecture, application architecture, and technology architecture.

Second, the *SAM* in [8, 9, 12, 14, 16-22, 25-29] in which the aligned domains are: Business Strategy, IT strategy, Business Infrastructure, IT infrastructure.

To standardize and simplify the evaluation of the literature, we adopted the SAM, which is described in Figure 1a, as the base of characterization.

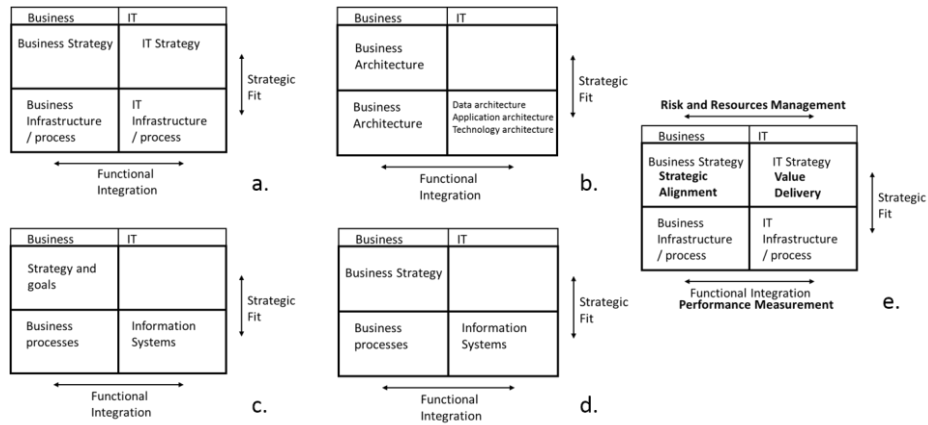

**Figure 1 a.** SAM domains [6]; **b.** EA domains mapped into SAM domains;
**c.** L.-H. Thevenet proposed alignment mapped into SAM; **d.** Doumi Et al. Proposed alignment model mapped into SAM; **e.** SAM and COBIT (Bold) domains combined

According to our decision of relying on the SAM as reference framework, we map in the Figure 1b the common EA domains used by the reviewed works, such as the domains of the EA Planning (EAP) [30], the layers of TOGAF [31] and the Zachman's levels [32]. Those are the most widespread EA models according to [33].

Differing from these common models, we have found some variants in terminology in the reviewed articles that we have mapped into the SAM domains. Thus, the Figure 1c shows the mapping of the Thevenet's approach [23] and the Figure 1d the mapping of the Doumi's approach [24]. In other custom cases we found that the SAM is combined with other frameworks. For instance, [18] presents a mix of the SAM and COBIT which we have mapped in the Figure 1e.

In this way we found that in BITA measurement, Business Infrastructure and IT infrastructure has been the focus in the 86% of the articles. The IT and Business strategies are only considered as main evaluated domains in 50% and 68% of the reviewed articles, respectively. This last finding reveals a lack of focus in the strategic layer.

**According to the SAM, which is the alignment sequence of the applied model?**

In order to describe the relationship or alignment between domains in which measurement is addressed in the reviewed works, we use alignment sequences of the SAM [6]. To this end, we use the domain roles used in [5] and the alignment perspectives proposed in [34] (See Table 2). The following domain roles are thus considered:

*Anchor:* the starting domain of the alignment sequence, represented by a square.

*Pivot:* the intermediate domain involved in the alignment sequence, represented by a circle.

*Impacted:* the final point of the alignment sequence, represented by an arrow head.

**Table 2.** Alignment Sequence Classification

| Alignment perspective | 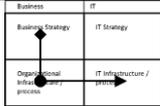 Figure 2 | 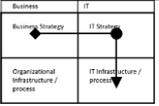 Figure 3 | 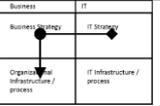 Figure 4 | 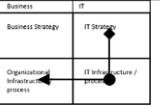 Figure 5 | 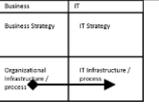 Figure 6 |
|---|---|---|---|---|---|
| Quantity of articles | 13 | 2 | 1 | 1 | 3 |
| Articles where sequence is addressed | [8, 11, 12, 14-20, 22, 23, 24] | [11, 12] | [10] | [21] | [9, 13, 25] |

The following alignment perspectives are considered:

*Strategy execution perspective.* It refers to a top-down sequence where the IT enables the business infrastructure and processes to execute the business strategies (see Figure 2).

*Technology potential.* In this perspective the focus is on the fit between business and IT strategies to enable new business strategy based on the technology application (see Figure 3).

*Competitive potential.* This perspective the focus is on the fit between business and IT strategies to do more competitive the business processes (see Figure 4).

*Service level.* This perspective focuses on deliver IT services and resources to enable business processes (see Figure 4).

*Functional integration.* This sequence focuses on the integrations of the business processes with the IT services and resources without an IT strategy (see Figure 6).

Considering the elements defined above for the evaluation of alignment sequences, the Table 2 show the quantity of research works that correspond better to each alignment perspective as well as the reference to those works. The results shows that the sequence in which alignment measurement is most carried out is Strategy Execution. This reflects that the traditional alignment sequence is the most addressed in the reviewed measurement approaches. The remaining sequences have a low development in the reviewed works.

Last, two of the reviewed articles [26, 27] are not included in the alignment sequence evaluation, because these are focused on developing frameworks to classify the current evaluation methods. Also, the works [28, 29] have not been included because they address alignment measurement in all the sequences, due to they are a SAMM [28] application cases.

**Which external environment elements of the business and IT are aligned?**

In this question, we use three external elements as possible responses [35]:

*Actors:* It refers to the external networks of actors present in the environment where organizations are involved.
*Uses:* It refers to the demands of the external environment of the organization, specifically referring to IT uses.
*Issues:* It groups all the external problematic situations where the organization could be involved and the barriers for the achievement of the objectives.

In the review, all the measurement methods and approaches focus only on the classical-internal level and, due to this, there are no external elements considered in the assessments. There is therefore a lack of external environment alignment measurement methods. This constitutes one of the main gaps that must be approached in future BITA measurement research.

**Which temporal dimensions are approached by the alignment model used?**
To answer this question, we have adopted two temporal states for the organizations [35]:

*As-Is:* referring to the current state.
*To-Be:* referring to the expected future state.

In this order, in all the reviewed work, the current state of the organization or *As-Is* is assessed. This show that measuring alignment of the current organizational and IT elements is the main objective of the BITA measurement processes. Even though the explicit assessment of the future state or *To-Be* is presented as the base for performance improvement, only three works [8, 20, 21] address alignment at this time-pitch. For instance, a fuzzy logic method is used in [8] to predict the future state based on the *As-Is* and the alignment sequence. The contributions of these three works are important because one of the main BITA measurement objectives is to establish a path for future action, but in general most of the methods only gives a diagnosis at the As-Is state.

**Which alignment levels are addressed by the work?**

In this review we found that in the 100% of the BITA measurement approaches, the classical-internal alignment level is the focus, only one approach of BITA measurement in [8] is related with the future state through a fuzzy logic expert system and there is no explicit focus on the external environment alignment. The articles [20, 21] mention the To-Be as desired state based on predictions but it is not really addressed and not is the main topic of the papers.

### 4.3 Measurement

**Which is the type of measurement method approached?**
According to [36] we adopted two options as possible response for the measurement type:

The *model-based* type which is focused on assessing the strength or quality of the link or relationship between the modelled elements in business and IT domains. Generally, the information used to carry out such measurement is obtained from documented models in the organization.

The *perception-based* type in which the evaluation is made from the perception of the actors (users, managers, etc.) in the different domains and levels of the organization. Generally, the information to undertake this type of measurement is obtained from surveys and interviews.

In our review we have found that approximately half of the reviewed works have each of these types of measurement, for model-based [8-10, 12, 13, 21, 23, 24, 26, 27] and for perception-based [11, 14-20, 22, 25, 28, 29]. It is logical, keeping in mind that the Luftman's method [28] (perception-based oriented) and the EA frameworks (model-based oriented) are the most widely adopted in the business environment.

Some of the last works like in [12], have begun to combine both perception and model-based measurement methods. This seems to be a more complex but also a more complete measurement approach.

**Which is the nature of the measurement method approached?**

To evaluate the nature of the measurement method, we define two classification categories, which are linked to the type criterion in last question, as possible response:

*Qualitative nature:* in this category the methods that yield a result are based on the quality of the alignment. These are not based on exact quantifiable scores that measure the alignment level. Conversely, they are mostly based on the perceptions of the quality of the fit between business and IT elements.

*Quantitative nature:* in this category the methods yield quantifiable scores based on the relationship between the modelled elements in the business and IT domains.

When reviewing the works, we found that the nature of the measurement methods used and developed are more oriented toward qualitative approaches. The results show that 50% of the articles have a qualitative nature and are based on evaluation scales that define a quality level of the alignment in the organizations [10, 11, 16, 17, 19, 20, 22, 25, 26, 28, 29], but with a wide range of error. It could be due to the high ambiguity of

9the comprehensive vision of the alignment in an enterprise scenario. The remaining reviewed articles are distributed in quantitative approaches in [9, 13, 15, 21, 23, 24, 27] (32%) and a mix of quantitative/qualitative approaches in [8, 12, 14, 18] (18%). The mixed nature could address a diagnosis with a more exact classification of the alignment levels, including a wide range of aspects that are defined as not only technically oriented.

**Which are the measurement criteria and metrics of the method?**

In the measurement process a variety of measurement criteria and metrics are defined, which are necessary to quantify or qualify the level of BITA. These criteria and metrics are closely related to the nature of the measurement method as the metrics define such nature. In the reviewed literature, the most adopted criteria by the reviewed articles (55%) correspond the SAMM six criteria, proposed by Luftman [28], which are used to measure the maturity of the BITA. The SAMM criteria are:

1. Communication
2. Competency/Value Measurements
3. Governance
4. Partnership
5. Technology Scope
6. Skills

Each one of these criteria are evaluated by using a five-level scale, in which each level possesses a list of defined best practices to be assessed in the organization. In some cases, the SAMM is completely adopted [28, 29], but, in general, the developed methods customize the five evaluation levels and the formulation of best practices as in [22] by using the Likert scale. Such changes are made in order to find a better match with the specific organizational departments and their culture. In this way, they obtain more accurate results in the assessment application.

In the remaining articles, in second position, some libraries of EA misalignment symptoms are included as the criteria and metrics base. These libraries are the comparative base to obtain quantifiable rates with the EA modelled elements in the organization. Finally, in our review we found that, in a lower proportion, some methods have custom metrics based on the experts' definitions and perceptions and in other modelling elements, like the ontologies, where the metrics are the rate of mapped ontologies from each domain in the organization. In these remaining metrics we also found: Key Performance Indicators (KPIs), COBIT control objectives and Crucial success factors (CSF).

**Which are the steps included in methodology applied?**

In this compilation of BITA measurement literature it is difficult to find a common structure of application methodology. This difficulty is due to the wide variety and heterogeneity of the current methods. Nevertheless, we found three general steps that are fundamental in any BITA measurement method. These steps are:

1. Describe the organization context by using modelling tools to have a parametrized scenario to work with.
2. Customize the BITA measurement tools to the context, organization departments and culture.
3. Assess all the organization departments.

## 5      Conclusions

The first conclusion of this literature review is that, in general, the measurement of the classical-internal BITA has been widely developed, supported in the need of performance improvement in the functioning of the organizations, at the level of internal processes and IT resources use. However, there is a lack of focus in BITA on the external environment of the companies. This is an important finding due to the current fast changing markets where traditional and modern companies are converging and competing. In addition, the new managing models such as outsourcing, offshoring and join-ventures in which alignment with external actors is indispensable has also increased the need for BITA measurement with the external extended organization.

The external BITA measurement could be addressed by including in the models and methodologies some frameworks such as Porter's five forces, presented in [37], which can help to complement the scope of the current methods and models.

Despite the fact that the SAM and the EA Frameworks are widely adopted, the selection of adequate tools and guidelines for BITA and its measurement continues to be complex due to the high ambiguity that BITA involves. The current advances are the basis for possible first steps toward a common standard for BITA, producing mixes of the methods, and producing more complete frameworks that are also contextualized with the organizations. The potential gap here is to find a common BITA standard which could be dynamic and flexible enough for the variety of organization contexts and approaches.

12